\documentclass[runningheads]{llncs}
\usepackage{graphicx}
\usepackage{hyperref}
\usepackage{xcolor}
\renewcommand\UrlFont{\color{blue}\rmfamily}
\usepackage{amsmath}
\usepackage{multirow}
\usepackage{bbm}

\begin{document}
\title{Physics-Informed Deep Learning for \\Motion-Corrected Reconstruction \\of Quantitative Brain MRI}
\titlerunning{PHIMO - Accepted at MICCAI 2024}

\author{Hannah Eichhorn\inst{1,2,*}
\and Veronika Spieker\inst{1,2}
\and Kerstin Hammernik\inst{2}
\and Elisa Saks\inst{3}
\and Kilian Weiss\inst{4}
\and Christine Preibisch\inst{3}
\and Julia A. Schnabel\inst{1,2,3,5}
}

\authorrunning{H. Eichhorn et al.}
\institute{Institute of Machine Learning in Biomedical Imaging, Helmholtz Munich, Germany
\and School of Computation, Information \& Technology, TUM, Germany
\and School of Medicine \& Health, TUM, Germany
\and Philips GmbH Market DACH, Germany
\and School of Biomedical Engineering \& Imaging Sciences, King’s College London, UK \\
*\email{hannah.eichhorn@tum.de}}

\maketitle
\begin{abstract}
We propose PHIMO, a physics-informed learning-based motion correction method tailored to quantitative MRI. PHIMO leverages information from the signal evolution to exclude motion-corrupted \mbox{k-space} lines from a data-consistent reconstruction. We demonstrate the potential of PHIMO for the application of T2* quantification from gradient echo MRI, which is particularly sensitive to motion due to its sensitivity to magnetic field inhomogeneities. 
A state-of-the-art technique for motion correction requires redundant acquisition of the \mbox{k-space} center, prolonging the acquisition.
We show that PHIMO can detect and exclude intra-scan motion events and, thus, correct for severe motion artifacts. PHIMO approaches the performance of the state-of-the-art motion correction method, while substantially reducing the acquisition time by over 40\%, facilitating clinical applicability.
Our code is available at \url{https://github.com/HannahEichhorn/PHIMO}.

\end{abstract}

% Statement of Novelty/Impact (max. 500 characters): 
% We introduce PHIMO, a physics-informed learning-based motion correction method designed for quantitative MRI, which utilises information from the signal evolution to exclude motion-corrupted k-space lines from a data-consistent reconstruction. For the application of T2* quantification from GRE MRI, we demonstrate that PHIMO achieves competitive image quality and significantly reduces the total acquisition time by over 40%, compared to a state-of-the-art motion correction method.

%
%
%
%%%%%%%%%%%%%%%%%%%%%%%%
% Introduction
%%%%%%%%%%%%%%%%%%%%%%%%
\section{Introduction}
Quantitative magnetic resonance imaging (MRI) estimates physical tissue properties from a series of qualitative images with varying imaging parameters. In contrast to commonly employed qualitative structural imaging, this facilitates a consistent extraction of potential biomarkers across scanners and hospitals. With typical voxel sizes of 2-3~mm, the resolution of quantitative MRI is commonly lower than for qualitative MRI due to longer acquisition times.

Patient motion is a challenge for brain MRI in general, potentially impeding successful diagnoses. Quantitative MRI is particularly sensitive to motion due to its intrinsically long imaging times. In recent years, deep learning-based approaches have demonstrated promising results for motion correction (MoCo) of brain MRI \cite{Spieker_2023}. MoCo has  previously been addressed as an image denoising problem, with convolutional \cite{Chatterjee_2020,Xu_2022} or generative adversarial networks \cite{Johnson_2019,Kustner_2019,Oh_2021}. Yet, exclusively relying on image data, these methods cannot ensure consistency with the acquired raw k-space data, posing a potential obstacle to their clinical translation. Data consistency (DC) is only achievable by integrating MoCo in the image reconstruction process \cite{Haskell_2019,Hossbach_2022,Singh_2023}. Nevertheless, previous methods have mostly been developed for higher resolution qualitative MRI or - in the context of quantitative MRI - they do not enforce DC \cite{Xu_2022}.

In this work, we propose physics-informed motion correction (PHIMO), which utilises information from the quantitative parameter estimation process to detect and exclude motion-corrupted k-space lines from a data-consistent reconstruction. We demonstrate the potential of PHIMO for the application of T2* quantification from gradient echo (GRE) MRI, which enables oxygenation-sensitive imaging as part of the multi-parametric quantitative BOLD (mqBOLD) protocol \cite{Hirsch_2014}. GRE MRI is highly sensitive to motion due to the influence of magnetic field inhomogeneities, especially for larger echo times \cite{Magerkurth_2011}. The current mqBOLD MoCo method \cite{Noth_2014} relies on redundant k-space acquisition, which significantly increases the total acquisition time from 3~min 39~s to 6~min 25~s. 

To the best of our knowledge, this is the first MoCo approach that incorporates information from the quantitative parameter estimation process as a physics-informed loss.
Our contributions are three-fold:
\begin{enumerate}
    \item We train a single unrolled reconstruction network to recover high-quality images from undersampled k-space data for varying rates of excluded lines.
    \item We train a multi-layer perceptron (MLP) to predict exclusion masks for motion-corrupted lines. Utilising the T2* decay information allows us to perform MoCo in a self-supervised fashion for each subject individually.
    \item We evaluate PHIMO on multicoil raw data acquired with and without head motion and compare the results to the current mqBOLD MoCo which requires redundant data acquisition.
\end{enumerate}

%%%%%%%%%%%%%%%%%%%%%%%%
% Background
%%%%%%%%%%%%%%%%%%%%%%%%
\section{Background}
\subsection{Motion during MR Image Acquisition} The MRI multicoil forward model in the presence of motion includes the sampling mask $\mathbf{S}_t$, incorporating the line-wise k-space acquisition pattern, the Fourier transform $\mathcal{F}$, the coil sensitivity maps $\mathbf{C}$, and the motion transform $\mathbf{U_t}$, which are applied to the  motion-free image $x$ for each time point $t$, yielding the motion-affected k-space data $\hat{y}$ \cite{Atkinson_2023}:
\begin{equation}
\label{eq:MRForward}
    \hat{y} = \sum_{t=1}^{T} \mathbf{S}_t \mathcal{F} \mathbf{C} \mathbf{U_t} x
\end{equation}
Head motion is typically treated as random rigid-body motion due to carelessness or discomfort. Consequently, $\mathbf{U_t}$ consists of rotation and translation transforms, but can also incorporate second-order motion effects, such as motion-induced field inhomgeneity changes \cite{Eichhorn_2023}. Under the assumption that the patient approximately stays in one position except for individual motion events, an undersampling mask $\mathbf{E}$ exists, which excludes the individual motion events, so that the masked motion-corrupted data is close to the masked motion-free data $y$:
\begin{equation}
    \mathbf{E} \hat{y} \simeq \mathbf{E} y
\label{eq:excl-motion}
\end{equation}

\subsection{T2* Quantification from Multi-Echo GRE MRI} The effective transverse relaxation time T2* can be quantified from a series of GRE images $x = [x_1, ..., x_E]$ acquired at various echo times~$t_e$ for $e=1, ..., E$. The time evolution of the signal magnitude $s_e = \lvert x_e \rvert$ for a single voxel is commonly modelled as a mono-exponential decay \cite{Magerkurth_2011}: 
\begin{equation}
    s_e = s_0 \cdot \exp\left(-\frac{t_e}{T2^*}\right)
    \label{eq:decay}
\end{equation}
T2* and the signal magnitude $s_0$ at $t_e=0$ follow from least-squares fitting.

%%%%%%%%%%%%%%%%%%%%%%%%
% Methods
%%%%%%%%%%%%%%%%%%%%%%%%
\section{Methods}
\subsection{PHIMO}
\begin{figure}[h!]
    \centerline{\includegraphics[width=\linewidth]{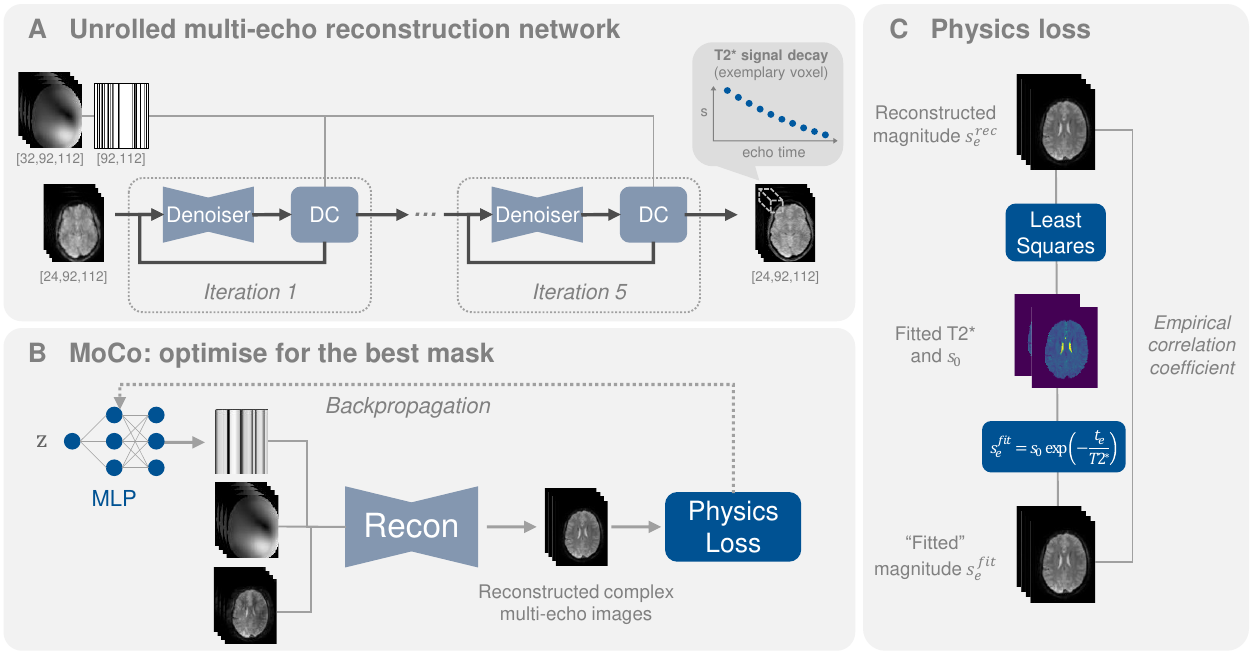}}
    \caption{Overview of PHIMO. (A) Training an unrolled reconstruction network on randomly undersampled \textit{motion-free} multi-echo images. (B) MoCo of \textit{motion-corrupted} data by optimising an MLP to predict slice-wise exclusion masks in a self-supervised fashion for each subject individually. (C) Calculation of empirical correlation coefficient as physics-informed loss for optimising the MLP in (B).}
    \label{fig:PHIMO-Overview}
\end{figure}

Inspired by MoCo through outlier-rejecting bootstrap aggregation \cite{Oh_2021}, we use Eq.~\ref{eq:excl-motion} to split MoCo into two subproblems: undersampled reconstruction and, in contrast to \cite{Oh_2021}, a physics-informed detection of motion-corrupted k-space lines. Specifically, Eq.~\ref{eq:excl-motion} allows us to train a reconstruction network on randomly undersampled \textit{motion-free} data $\mathbf{E}y$ and, during inference, recover a high-quality image from the undersampled \textit{motion-corrupted} data $\mathbf{E}\hat{y}$, given that $\mathbf{E}$ excludes the motion-corrupted k-space lines. An overview of PHIMO is provided in Fig.~\ref{fig:PHIMO-Overview}. 

\subsubsection{Supervised Training of Unrolled Reconstruction}
First, we train an unrolled multi-echo reconstruction network on randomly undersampled \textit{motion-free} data.
The reconstruction alternates five times between a CNN-based denoiser and a gradient descent DC step (see Fig. \ref{fig:PHIMO-Overview}A), without weight sharing between iterations. The denoiser consists of five 2D convolutional layers (kernel size: $3\times3$, 64 filters) and ReLU activations.
Real and imaginary parts of the input image, as well as multiple echoes are stacked in the channel dimension. The network is trained for 4000 epochs with Adam~\cite{Adam_2015} and a batch size of eight. We use a learning rate of $1\times10^{-4}$ and structural similarity (SSIM) \cite{Wang_2004} as loss, calculated on real and imaginary parts separately. During training, the rate of excluded lines in the undersampling masks is varied between 0.05 and 0.5, with a fixed fully-sampled center of 10 lines.

\subsubsection{Self-Supervised k-Space Line Detection}
In the presence of \textit{motion-cor-rupted} data, we perform MoCo by detecting and excluding motion-corrupted k-space lines from the reconstruction.
Therefore, we train an MLP for each subject to predict an undersampling mask that excludes motion-corrupted k-space lines (Fig. \ref{fig:PHIMO-Overview}B). We introduce a physics loss, which takes into account that for multi-echo GRE MRI, motion increasingly impacts the image with increasing echo time. Thus, motion disturbs the mono-exponential signal evolution (Eq.~\ref{eq:decay}) \cite{Noth_2014}.
As illustrated in Fig.~\ref{fig:PHIMO-Overview}C, the physics loss compares the reconstructed signal intensities $s_e^{rec}$ and the “fitted” intensities $s_e^{fit}$ resulting from inserting the fitted parameters T2* and $s_0$ into Eq.~\ref{eq:decay}. We calculate the physics loss $L_{phys}$ as the empirical correlation coefficient, with the mean over all echoes $\overline{s} = \frac{1}{E} \sum_{e} s_e$: 
\begin{equation}
    L_{phys} = \frac{\sum_e (s_e^{rec} - \overline{s}^{rec}) (s_e^{fit} - \overline{s}^{fit})}
    {\sqrt{\sum_e (s_e^{rec} - \overline{s}^{rec})^2 (s_e^{fit} - \overline{s}^{fit}})^2}
\end{equation}
$L_{phys}$ is calculated voxel-wise across echoes and averaged within a brain mask excluding cerebrospinal  fluid (CSF). We combine $L_{phys}$ with a regularisation~$L_{reg}$ on the variation of the predicted masks for adjacent slices. Following the interleaved, multi-slice acquisition scheme, $L_{reg}$ is calculated between all $Z$ pairs of predicted masks, $\mathbf{E}_z$ and $\mathbf{E}_{z+2}$, within the batch:
\begin{equation}
     L = L_{phys} + \lambda L_{reg} = L_{phys} 
     + \lambda \sum_{z=1}^{Z} (\mathbf{E}_z - \mathbf{E}_{z+2})
\end{equation}
The MLP embeds the slice index $z$ using a mapping function $f \colon \mathbbm{R}^1 \rightarrow \mathbbm{R}^3$, % into a vector of size 3
followed by two fully connected layers of size 23 and 46 and a sigmoid activation. It results in an output mask of size 92, with values ranging between 0 and 1. We do not binarise the output mask to allow for a smooth loss minimisation. Note that we also do not restrict the output masks to a fully-sampled center to allow for an exclusion of motion-corrupted lines in the k-space center.
The MLP is optimised with Adam, a batch size of 4 and a regularisation strength of $\lambda=0.1$. We perform early stopping when $L_{reg}$ does not improve for 50 epochs. Optimising the MLP takes less than 14 minutes per subject.

\subsection{Data}
We have acquired multi-coil k-space data from 15 volunteers (26.6 ± 2.8 years, 5 females) on a 3T Philips Elition X MR scanner (Philips Healthcare, Best, The Netherlands), using a multi-slice 2D GRE sequence (12 echoes, TE1=$\Delta$TE=5~ms, TR=2300~ms, voxel size: 2$\times$2$\times$3~mm, 32-channel head coil). The study has been approved by the local ethics committee (approval numbers 440/18 S-AS, 2023-386-S-SB).
We have conducted repeated scans under two conditions: without voluntary head motion and with the subject instructed to move randomly, imitating sneezing or coughing. In one case, referred to as \textit{motion timing experiment}, we have provided precise timings for the volunteer to move, aiming to assess the motion detection performance of PHIMO. 
Additionally, we have acquired half- and quarter-resolution data in both conditions to compare PHIMO to the current mqBOLD MoCo (“HR/QR-MoCo”) \cite{Noth_2014}. The datasets are divided subject-wise into train, validation, and test sets (6/2/7 subjects), including only slices with more than 10\% brain voxels (193/61 train/validation slices). For testing, we manually exclude inferior slices to disentangle motion from severe susceptibility artifacts, which need to be  addressed separately \cite{Hirsch_2013}, resulting in 128 test slices.
Prior to the reconstruction, the data are normalised by the maximum image magnitude. 

\subsection{Evaluation}
We compare PHIMO to the state-of-the-art HR/QR-MoCo \cite{Noth_2014} and outlier-rejecting bootstrap aggregation (OR-BA) \cite{Oh_2021}. The latter averages reconstructions with 15 random masks, for which we train a separate unrolled reconstruction network with variable density masks at a fixed exclusion rate of 0.5. To differentiate between the methods' performances for more severe and minor motion, we categorise the seven test subjects based on the visual quality of the motion-corrupted images, resulting in four subjects with severe and three with minor motion.
T2* maps are evaluated based on mean absolute error (MAE), SSIM and feature similarity (FSIM)~\cite{Zhang_2011}. Therefore, all acquisitions are aligned to the motion-free acquisition via 3D registration of the stacked slices. Wilcoxon signed rank tests and False-Discovery Rate correction are employed for statistical testing. 
HR/QR-MoCo and segmentation of anatomical scans are performed in MATLAB (R2022b) and SPM12 with custom programs \cite{Kaczmarz_2020b}. Other computations are performed in Python 3.8.12, using PyTorch 2.0.1 and MERLIN~\cite{HammernikKuestner2022}.

%%%%%%%%%%%%%%%%%%%%%%%%
% Results
%%%%%%%%%%%%%%%%%%%%%%%%
\section{Experiments and Results} 
\begin{figure}[b]
    \centerline{\includegraphics[width=\linewidth]{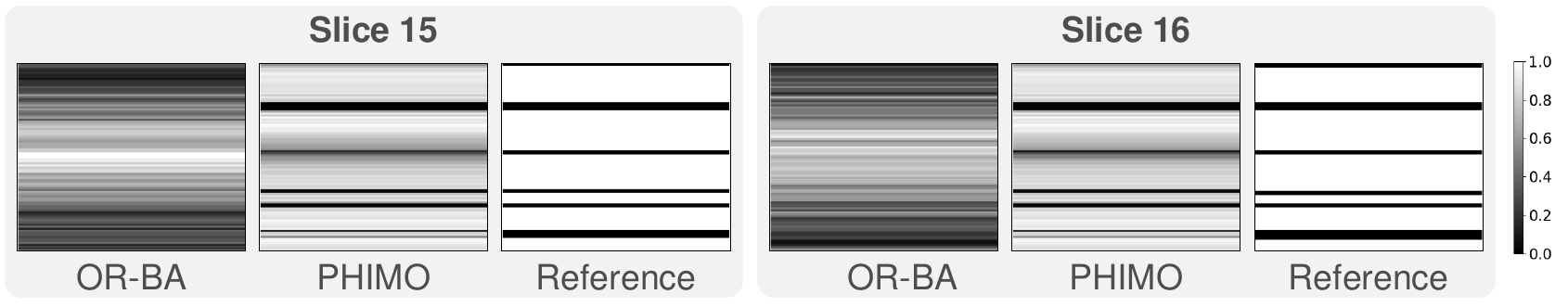}}
    \caption{Exclusion masks of two example slices for the motion timing experiment (single subject), derived from OR-BA, PHIMO and the verbal instructions (reference). The corresponding reconstructions with reduced motion artefacts for PHIMO are provided in the supplementary material (Fig. S1).}
    \label{fig:Example_Masks}
\end{figure}
In Fig.~\ref{fig:Example_Masks} we compare exclusion masks derived from OR-BA and PHIMO with the reference mask inferred from the verbal instructions of the motion timing experiment. This example demonstrates that PHIMO can detect individual motion events. In contrast, OR-BA on average assigns nearly uniform weights to all k-space lines and cannot detect specific lines.
\begin{figure}[h!]
    \centerline{\includegraphics[width=\linewidth]{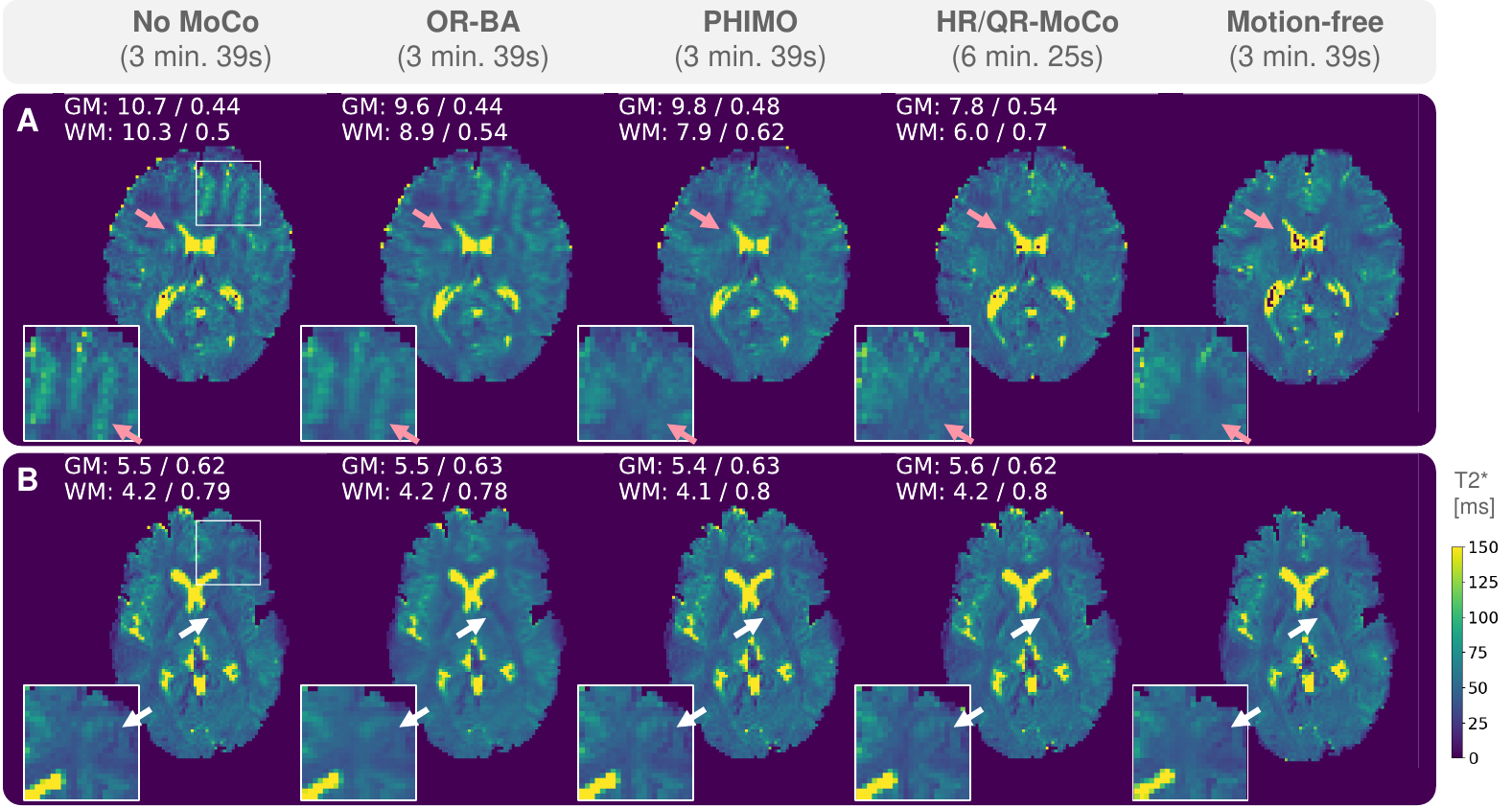}}
    \caption{Example T2* maps for subjects with severe (A) and minor motion (B). From left to right: motion-corrupted (No MoCo), OR-BA, proposed PHIMO, HR/QR-MoCo and motion-free reference. Pink arrows indicate (suppressed) motion artifacts, white arrows increased blurring for OR-BA. Gray/white matter (GM/WM) MAE values in ms as well as SSIM values relative to the motion-free maps are indicated in the top left corner. The corresponding exclusion masks of OR-BA and PHIMO are presented in the supplementary material (Fig. S2).}
    \label{fig:Example_T2star}
\end{figure}
Example T2* maps in Fig.~\ref{fig:Example_T2star} demonstrate that - in contrast to OR-BA - PHIMO effectively suppresses motion artefacts for the subject with severe motion, close to the performance of HR/QR-MoCo. In the case of minor motion, OR-BA leads to blurring, while PHIMO preserves the image quality and is comparable to the motion-free reference. Note that, despite the volunteer's cooperation, residual motion may still be present in the so-called motion-free reference acquisition.
\begin{figure}
    \centerline{\includegraphics[width=\linewidth]{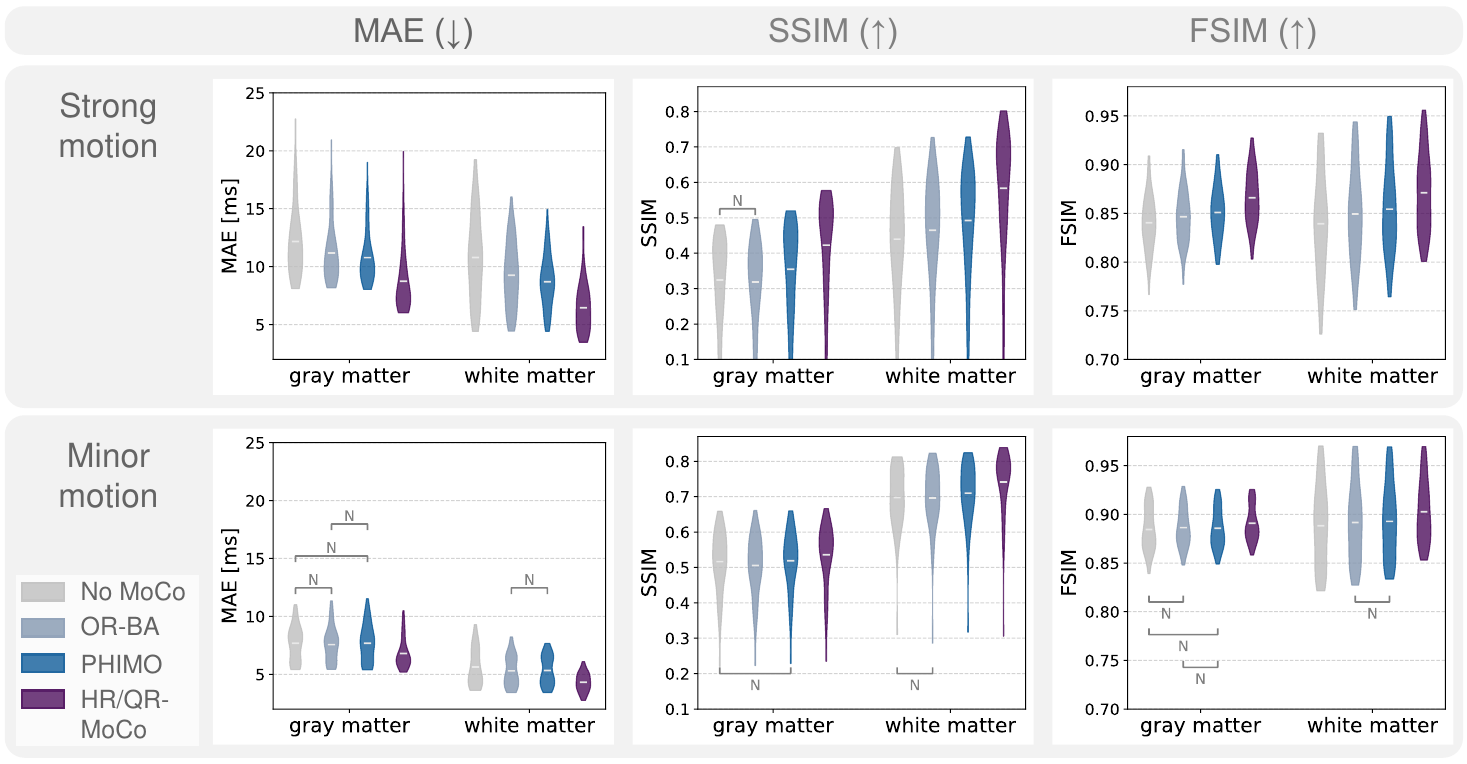}}
    \caption{Image quality metrics MAE, SSIM and FSIM, calculated relative to the motion-free T2* maps, for the 4 test subjects with severe (top) and the 3 test subjects with minor motion (bottom). All metrics are calculated on T2* maps registered to the motion-free reference and evaluated in gray and white matter, separately. Gray brackets indicate comparisons with no statistical significance (N) at a level of $p=0.001$.}
    \label{fig:Quant_T2star}
\end{figure}
The qualitative observations are supported by the quantitative analysis in Fig.~\ref{fig:Quant_T2star}. In case of more severe motion, all metrics are notably improved for PHIMO, approaching the values of HR/QR-MoCo. OR-BA yields inconsistent results: for MAE it results in significantly but not substantially worse values than PHIMO, while PHIMO more clearly outperforms OR-BA for the structure-sensitive metrics SSIM and FSIM. 
For minor motion, the values for the motion-corrupted maps are already satisfactory, but PHIMO and HR/QR-MoCo achieve small improvements, particularly in white matter regions. The performance of OR-BA remains inconsistent across the evaluated metrics.

%%%%%%%%%%%%%%%%%%%%%%%%
% Discussion
%%%%%%%%%%%%%%%%%%%%%%%%
\section{Discussion and Conclusion}
We propose PHIMO, a physics-informed MoCo method tailored to quantitative MRI. To the best of our knowledge, PHIMO is the first approach that incorporates information from the signal evolution in order to down-weight motion-corrupted k-space lines in a DC-based reconstruction. For the application of T2* quantification from multi-echo GRE MRI, we have demonstrated that PHIMO can detect motion events, suppress strong motion artefacts and thus, significantly improve the quality of motion-corrupted T2* maps. 

To investigate the impact of our proposed physics-informed loss for excluding motion events and satisfying equation \ref{eq:excl-motion}, we have compared PHIMO to OR-BA, which we have implemented as a bootstrap aggregation averaging reconstructions with random masks, in line with the idea of Oh et al. \cite{Oh_2021}. The visual examples in Fig.~\ref{fig:Example_T2star} and the structure- and feature-based quantitative metrics SSIM and FSIM in Fig.~\ref{fig:Quant_T2star} demonstrate that PHIMO notably outperforms OR-BA. A reason for that is that OR-BA samples random masks and, thus, down-weights all k-space lines uniformly on average (Fig.~\ref{fig:Example_Masks}), thereby lacking the ability to adequately detect and exclude corrupted lines. 

MAE does not seem to reflect PHIMO's - visually apparent - superiority to OR-BA in suppressing strong motion artefacts. In general, reliable image quality evaluation is an ongoing challenge in the MoCo and image reconstruction community \cite{Spieker_2023}. In our case, MAE's inconsistency could be due to PHIMO excluding central k-space lines in the case of detected motion events, which allows for a reduction of strong motion artefacts, as highlighted in Fig.~\ref{fig:Example_T2star}. 
At the same time, excluding central k-space lines might lead to some T2* estimation errors, since the k-space center contains overall signal intensity and contrast information.
Nevertheless, it appears preferable to exclude severe motion outliers while slightly misestimating T2* (PHIMO) rather than including motion measurements and not correcting severe artefacts (OR-BA).  

We have further compared PHIMO to HR/QR-MoCo \cite{Noth_2014}, which is the state-of-the-art within the mqBOLD technique \cite{Kaczmarz_2020b}. HR/QR-MoCo combines the motion-corrupted scan with additionally acquired half- and quarter-resolution scans, effectively sampling the k-space center three times. 
The performance of PHIMO as well as HR/QR-MoCo depends on the exact motion pattern. Both approaches build on the assumption that the subject generally maintains a consistent position, interrupted by random motion events. Additionally, PHIMO is more severely impacted by subject motion that occurs during the acquisition of the k-space center, since excluding low-frequency contrast information affects the reconstruction quality more severely than missing high-frequency information on edges and details. 
In our qualitative and quantitative comparison, HR/QR-MoCo outperforms PHIMO for more severe and minor motion scenarios, as it essentially results in a weighted average of three acquisitions, which even for a motion-free scan leads to noise reduction. However, PHIMO achieves a competitive image quality, particularly when taking into account that it does not rely on additionally acquired half- and quarter-resolution scans. Consequently, PHIMO reduces the total acquisition time by over 40~\%, which is a key for facilitating clinical applicability of the entire mqBOLD technique.

\subsubsection*{Conclusion}
We have introduced PHIMO, a MoCo method tailored to quantitative MRI, which utilises information from the MR signal evolution to detect motion events with a subject-specific, self-supervised MLP.
We have demonstrated PHIMO's capabilities in detecting and adequately down-weighting motion-cor-rupted k-space lines in a DC-based reconstruction for T2* quantification from GRE MRI. Compared to the state-of-the-art motion correction method, which requires redundant data acquisition, PHIMO achieves competitive image quality and reduces the overall acquisition time from 6~min 25~s to 3~min 39~s, positively impacting clinical workflow. Importantly, PHIMO is applicable to various other quantitative MRI tasks, as long as the parameter estimation is differentiable.

%%%%%%%%%%%%%%%%%%%%%%%%
% Acknowledgements
%%%%%%%%%%%%%%%%%%%%%%%%
\begin{credits}
\subsubsection{\ackname} V.S. and H.E. are partially supported by the Helmholtz Association under the joint research school ”Munich School for Data Science - MUDS”.

\subsubsection{\discintname} K.W. is an employee of Philips GmbH Market DACH.
\end{credits}

% \newpage
\renewcommand\UrlFont{\color{black}\rmfamily}

\bibliographystyle{splncs04}
\bibliography{references}

\newpage

\section*{Supplementary Material}
\setcounter{figure}{0}
\renewcommand{\figurename}{Fig.}
\renewcommand{\thefigure}{S\arabic{figure}}

\begin{figure}[h!]
    \centering
    \includegraphics[width=\linewidth]{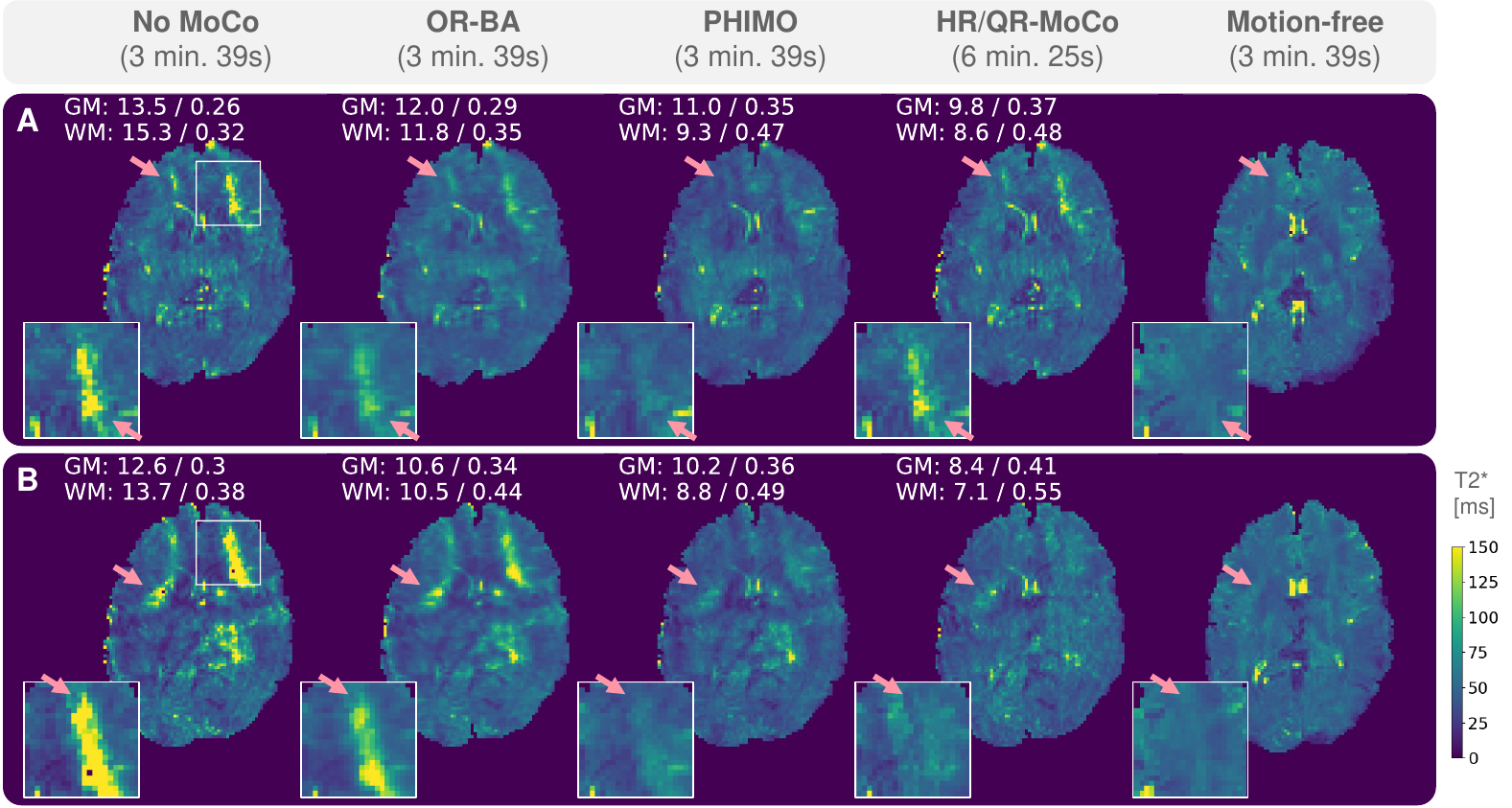}
    \caption{T2* maps corresponding to the exclusion masks in Fig.~2 of the main article, showing slice 15 (A) and slice 16 (B) of the subject performing the motion timing experiment. From left to right: motion-corrupted (No MoCo), OR-BA, proposed PHIMO, HR/QR-MoCo and motion-free reference. Pink arrows indicate (suppressed) motion artifacts. Gray/white matter (GM/WM) MAE values in ms as well as SSIM values relative to the motion-free maps are indicated in the top left corner.}
    \label{fig:Suppl_1}
\end{figure}

\begin{figure}[h!]
    \centerline{\includegraphics[width=\linewidth]{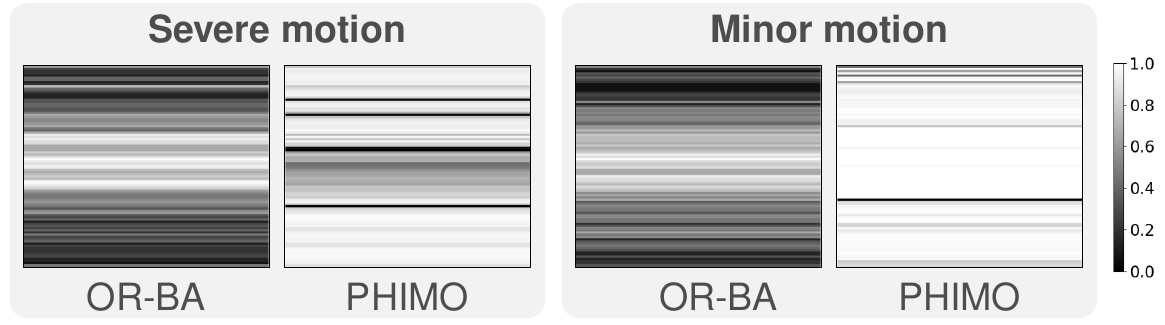}}
    \caption{Exclusion masks derived from OR-BA and PHIMO, corresponding to the example T2* maps in Fig.~3 of the main article for more severe motion (left) and minor motion (right). Note that these subjects were instructed to move at random time points during the acquisition and not given exact timing instructions like the example in Fig.~\ref{fig:Suppl_1}. Thus, no reference mask is available for these subjects.}
    \label{fig:Suppl_2}
\end{figure}

\end{document}